\documentclass[pdflatex,sn-mathphys-num]{sn-jnl}

\usepackage{graphicx}%
\usepackage{multirow}%
\usepackage{amsmath,amssymb,amsfonts}%
\usepackage{amsthm}%
\usepackage{mathrsfs}%
\usepackage[title]{appendix}%
\usepackage{xcolor}%
\usepackage{textcomp}%
\usepackage{manyfoot}%
\usepackage{booktabs}%
\usepackage{algorithm}%
\usepackage{algorithmicx}%
\usepackage{algpseudocode}%
\usepackage{listings}%

\usepackage{csquotes}
\MakeOuterQuote{"}

\usepackage{siunitx}

\usepackage{bm}
\usepackage{braket}
\usepackage{physics}

\renewcommand{\v}[1]{\bm{\mathrm{#1}}}
\newcommand{\m}[1]{\bm{\mathsf{#1}}}

\theoremstyle{thmstyleone}%
\theoremstyle{thmstyletwo}%

\theoremstyle{thmstylethree}%

\begin{document}

\title[Article Title]{Valley polarization of graphene via the saddle point}

\author[2]{\fnm{Deepika} \sur{Gill}}\email{deepikagill1008.dg@gmail.com}

\author[1,2]{\fnm{Sangeeta} \sur{Sharma}}\email{geet1729@gmail.com}

\author[3]{\fnm{Peter} \sur{Elliott}}\email{elliottp.physics@gmail.com}

\author[4]{\fnm{Kay} \sur{Dewhurst}}\email{dewhurst@mpi-halle.mpg.de}

\author*[4]{\fnm{Sam} \sur{Shallcross}}\email{phsss75@gmail.com}

\affil*[1]{\orgname{Max-Born-Institute for Non-Linear optics}, \orgaddress{\street{Max-Born Strasse 2A}, \city{Berlin}, \postcode{12489}, \country{Germany}}}

\affil[2]{\orgdiv{Institute for theoretical solid-state physics}, \orgname{Freie Universit\"at Berlin}, \orgaddress{\street{Arnimallee 14}, \city{Berlin}, \postcode{14195}, \country{Germany}}}

\affil[3]{\orgdiv{Scientific Computing Department}, \orgname{Science and Technology Facilities Council UK Research and Innovation (STFC-UKRI), Rutherford Appleton Laboratory}, \orgaddress{\city{Didcot}, \postcode{OX11 0QX}, \country{United Kingdom}}}

\affil[4]{\orgname{Max-Planck-Institut fur Mikrostrukturphysik}, \orgaddress{\street{Weinberg 2}, \city{Halle}, \postcode{D-06120}, \country{Germany}}}


\abstract{
Graphene, and other members of the monolayer Xene family, represent an ideal materials platform for "valleytronics", the control of valley localized charge excitations. The absence of a gap in these semi-metals, however, precludes valley excitation by circularly polarized light pulses, sharply circumscribing the possibility of a lightwave valleytronics in these materials. Here we show that combining a deep ultraviolet linearly polarized light pulse with a THz envelope can induce highly valley polarized states in graphene. This dual frequency lightform operates by (i) the deep ultraviolet pulse activating a selection rule at the M saddle points and (ii) the THz pulse displacing the M point excitation to one of the low-energy K valleys. Employing both tight-binding and state-of-the-art time dependent density functional theory, we show that such a pulse results in a near perfect valley polarized excitation in graphene, thus providing a route via the saddle point to a lightwave valleytronics in the gapless Xene family.
}

\maketitle

\section{Introduction}

Selection rules underpin a vast range of light-matter effects in the solid state, both during ultrafast non-equilibrium dynamics as well as in the linear response regime. This is exemplified by the light-valley selection rule of the $2H$ transition metal dichalcogenides (TMDC) that couples opposite helicity circularly polarized light to conjugate K and K$^\ast$ low energy valleys\cite{xiao_coupled_2012}, and that determines both the ultrafast and long time response to light of these materials. The controlled light activation of valley states that this allows represents one of the most promising routes towards the writing of information in quantum matter -- with 0 and 1 represented by charge excitation at conjugate valleys -- and underpins the emerging field of "valleytronics"\cite{
schaibley_valleytronics_2016,
higuchi_light-field-driven_2017,
vitale_valleytronics_2018,
langer_lightwave_2018,
heide_lightwave-controlled_2019,
mciver_light-induced_2020-1,
jimenez-galan_lightwave_2020,
sharma_valley_2022}.

Such light-valley coupling requires, however, a gapped electronic spectrum, excluding from this physics the family of Xenes: graphene, stanene, and silicene. Graphene, in particular, represents an ideal material for valleytronics as the low energy Dirac-Weyl transport physics entails both strong suppression of valley scattering as well as the possibility for control over valley currents by gauge field engineering
\cite{zhai_valley-filtering_2011,settnes_graphene_2016,
 gupta_straintronics_2019,
 hsu_nanoscale_2020}. Sustained attempts have thus been made to "bypass" this absence of a selection rule for light-valley coupling and generate, by light pulses, valley pure charge excitation in this material\cite{
mrudul_light-induced_2021,
mrudul_controlling_2021,
avetissian_graphene_2023,
kelardeh_ultrashort_2022}.

Valley trigonal warping -- a $C_3$ deformation of the band manifold  -- and the fact that conjugate valleys are related by mirror symmetry can be utilized to create a charge imbalance between the valleys
\cite{
mrudul_light-induced_2021,
mrudul_controlling_2021,
avetissian_graphene_2023,
kelardeh_ultrashort_2022}. However the excited states attained in this way achieve only 20-60\% valley polarization (defined as the normed difference of the valley charges) and, furthermore, represent states in which charge is excited quite widely over the Brillouin zone, i.e. states with significant non-valley charge. This situation is evidently very different from the lightwave control achieved in the TMDC family or gapped bilayer graphene\cite{friedlan_valley_2021}, in which strongly localized 100\% valley polarized excitations can be created without extraneous non-valley charge excitation.

Recently, Sharma {\it et al}. have proposed an alternative mechanism involving a dual frequency (double pumped) laser pulse consisting of infra-red circularly polarized and THz\cite{bera_review_2021,
mashkovich_terahertz_2021,
khan_ultrafast_2020,chekhov_ultrafast_2021} linearly polarized components\cite{sharma_combining_2025}. The first of these components excites -- in the absence of a selection rule in graphene -- charge at both the K and K$^\ast$ low energy valleys. This valley unpolarized excitation is then displaced in momentum space by the THz pulse component resulting, in principle, in a high energy $\Gamma$-point excitation along with a low energy K point excitation. Remarkably, however, a pronounced de-excitation occurs as charge evolves towards the high energy $\Gamma$-point, thus yielding a final state of 90-100\% valley polarization, with very low extraneous non-valley charge.

Here we explore an alterative scheme consisting of a double pumped pulse of two linearly polarized components: a deep ultraviolet and a THz component with orthogonal polarization. This, as we show, provides a route towards the creation of a valley localized excitation in graphene via the M saddle point of graphene. The ultraviolet component couples predominantly to one of the three inequivalent M points of graphene, with the charge excited at the M point saddle then "displaced" to a K valley via the THz pulse component. In comparison to valley polarization via the $\Gamma$ point discussed in Ref.~\cite{sharma_combining_2025} this has two advantages: (i) a reduced THz amplitude is required (as the M-K momentum separation is half that of the $\Gamma$-K momentum separation); and (ii) the laser pulse employs only linearly polarized components and thus there is no requirement for carrier envelope phase stability.

The remainder of this paper is structured as follows: in section 2 we establish a route to the polarization of graphene via the saddle point, in section 3 we explore how this polarization can be controlled by light pulse parameters, and in section 3 present {\it ab-initio} calculations of valley polarization in graphene. Finally, in section 4 we conclude and discuss the various pulse designs "in the market" for valley polarization of graphene .

\section{Simulation methods}

For the dynamical simulation of light-matter interaction in graphene will employ two complementary methodologies: (i) the tight-binding (TB) method and (ii) time dependent density functional theory (TDDFT). The former of these represents a highly numerically efficient approach, and can be used to explore and optimize pulse parameters. This, however, is at the price of a dynamics involving only the occupation numbers. The TDDFT approach, in contrast, involves dynamical evolution of the full density $\rho(\v r, t)$ and represents a "gold standard" for early time simulation of laser induced dynamics
\cite{dewhurst_efficient_2016,dewhurst_laser-induced_2018,siegrist_light-wave_2019}. 
We employ this method to verify our findings based on the tight-binding method.

The laser pulse form we take to be a Gaussian envelope centred at $t_0$ modulating a sinusoidal oscillation, and construct the total pulse as a sum of these:

\begin{equation}
\v A(t) = \sum_i \v A_0^{(i)} \exp\left(-\frac{(t-t_0^{(i)})^2}{2\sigma_i^2}\right) \sin\left[\omega_i (t-t_0^{(i)}) + \phi_i\right].
\label{pulse}
\end{equation}
Here $\v A_0^{(i)}$ represents the pulse component polarization vector, $\sigma_i$ is related to the full width half maximum by $FWHM = 2\sqrt{2\ln{2}}\sigma$, and $\omega_i$ the central frequency of the pulse component.

\subsection{Tight-binding simulations}

A minimal tight-binding model for graphene consists of the $\pi$-band only Hamiltonian

\begin{equation}
H_0 = \begin{pmatrix}
0 & t(\v k) \\ t(\v k)^\ast & 0
\end{pmatrix}
\label{H0}
\end{equation}
where $t(\v k) = -t \sum_j e^{i\v k.\m\nu_j}$ is the Bloch sum over nearest neighbours $\m\nu_j$, and $t=-2.8$~eV is the nearest neighbour hopping.

{\it Time propagation within the tight-binding approach}:  The initial state is provided by a Fermi-Dirac distribution with $T=0$, and we expand the time-dependent wavefunction at crystal momenta $\v q$ using Bloch states $\ket{\Phi_{\alpha \v k(t)}}$ where $\alpha=1,2$ labels the two sub-lattices of graphene:

\begin{equation}
\ket{\Psi_{\v q}(t)} = \sum_{\alpha} c_{\alpha \v q}(t) \ket{\Phi_{\alpha \v k(t)}}
\label{eq:S1}
\end{equation}
where $\v k(t)$ is given by the Bloch acceleration theorem

\begin{equation}
\v k(t) = \v q - \v A(t)/c
\end{equation}
The time-dependent Schr\"odinger equation can then be written in terms of the column vector of these time dependent expansion co-efficients $c_{\alpha \v q}(t)$ as

\begin{equation}
i \partial_t c_{\v q}(t) = H_0(\v k(t)) c_{\v q}(t)
\label{eq:cSE}
\end{equation}

For the numerical propagation of Eq.~\ref{eq:cSE} we employ the Crank-Nicolson method with a time step of 20 attoseconds. For variation of the pulse parameters of the excitation part of the pulse, a $300\times 300$ k-mesh is employed. Variation of the pulse parameters of the THz envelope pulse require significantly higher convergence, due to the complex interference patterns created, and a $2600\times 2600$ k-mesh is used in this case.

\subsection{Time density functional theory}

TD-DFT~\cite{runge1984,sharma2014} rigorously maps the computationally intractable problem of interacting electrons to a Kohn-Sham system of non-interacting electrons in an effective potential. The time-dependent KS equation is:
\begin{align}
\begin{split}
i \frac{\partial \psi_{j}({\bf r},t)}{\partial t} =
\Bigg[
\frac{1}{2}\big(-i{\nabla}&-\frac{1}{c}{\bf A}(t)\big)^2
+ v_{s}({\bf r},t) \Bigg]
\psi_{j}({\bf r},t),
\end{split}
\label{e:TDKS}
\end{align}
where $\psi_j$ is a KS orbital and the effective KS potential $v_{s}({\bf r},t) = v({\bf r},t)+v_{\rm H}({\bf r},t)+v_{\rm xc}({\bf r},t)$ consists of the external potential $v$, the classical electrostatic Hartree potential $v_{\rm H}$ and the exchange-correlation (XC) potential $v_{\rm xc}$. The vector potential ${\bf A}(t)$ represents the applied laser field within the dipole approximation (i.e., the spatial dependence of the vector potential is absent).

{\it Computational parameters for the TD-DFT calculations}: In our calculations we employ a $30\times 30 \times 1$ k-mesh. All states up to an energy of 60~eV above Fermi energy are included, and the adiabatic local density approximation (LDA) is used for $v_{xc}$. The time step is 2.4 attoseconds is used for time-propagation (for details of the time-propagation algorithm see Ref.~\cite{dewhurst_efficient_2016}). An in-plane lattice parameter of $a=b=2.42$~\si{\angstrom} was used and in order to simulate a mono-layer a vacuum of $20.0$~\si{\angstrom} was used in the $c$-direction.

\section{From saddle to valley}

For our pulse design we adopt a twofold strategy: (i) we first consider a light pulse generating localized excitation at one of the three inequivalent M points\cite{sharma_ultrafast_2025}, and then (ii) via intra-band motion shift this excitation to one of the two inequivalent K points. Our pulse thus consists of two components, one of which generates the light induced inter-band excitation at the M saddle, and one the light induced intra-band excitation from M saddle to K valley.

\begin{figure}[t!]
\centering\includegraphics[width=1.0\textwidth]{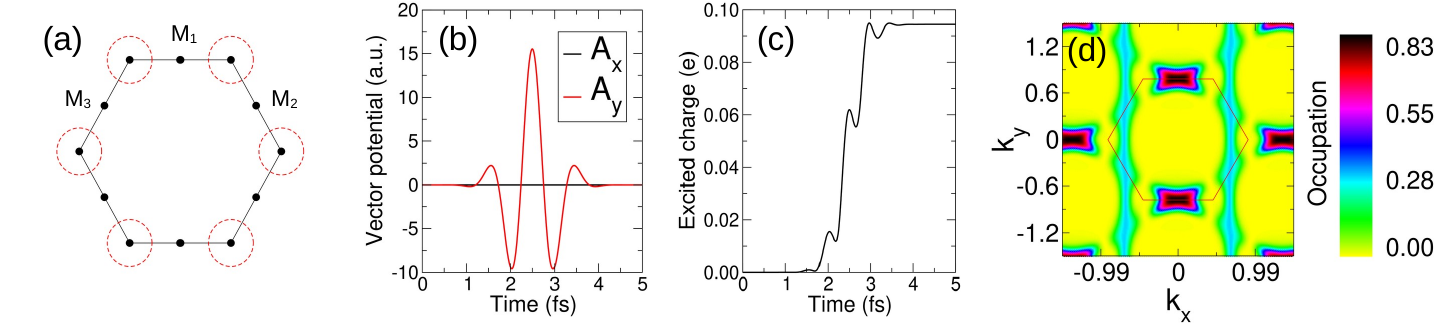}
\caption{\it{Saddle point polarization in graphene.} (a) The Brillouin zone of graphene with the three inequivalent M points labelled. A few cycle deep ultraviolet pulse tuned to the 4~eV M point gap, vector potential shown in panel (b), generates a charge excitation, panel (c). This charge excitation is strikingly inhomogeneous in momentum space, with the excitation dominant at one of the three inequivalent M points, panel (d).
}
\label{fig1}
\end{figure}

A selection rule for the M saddle point of graphene has recently been described by Sharma {\it et al}., and we here we present a brief description of this referring the reader to Ref.~\cite{sharma_ultrafast_2025} for full details.
In the velocity gauge, and employing the dipole approximation, the coupling of light to the graphene Hamiltonian Eq.~\ref{H0} is achieved via the vector potential $\v A(t)$ as

\begin{equation}
H(t) = H_0 + \m \nabla_{\v k} H_0(\v k).\v A(t)
\end{equation}
Taylor expansion of Eq.~\ref{H0} at the three inequivalent M points shows the velocity operator to be given by

\begin{equation}
\m \nabla_{\bf k} t(\v M_i) = -i\frac{\alpha a t^2}{\pi} \v M_i
\end{equation}
where $\v M_i$ represents the crystal momentum of one of the three inequivalent M points, see Fig.~\ref{fig1}(a), and $\alpha=2\pi/3$.

To establish light-matter coupling at the M points we consider a harmonic pulse $\v A = \v A_0 \cos\omega t$, for which the Fermi golden rule then gives for the excitation probability

\begin{equation}
T(\omega) = \frac{2 a^2 t^4}{\pi} (\v M_i.\v A_0)^2 \delta(\varepsilon_M - \omega)
\label{select}
\end{equation}
where $\varepsilon_M=4$~eV is the gap at the M point.
If the polarization vector $\v A_0$ is aligned parallel with one of the three inequivalent M points, say $M_1$, such that $\hat{\v M}_1.\hat{\v A}_0 = 1$ (with $\hat{\v M}_1$ and $\hat{\v A}_0$ the unit vectors corresponding to $\v M_1$ and $\v A_0$), then the remaining two M points, related by 60$^\circ$ rotation, will have $\hat{\v M}_i.\hat{\v A}_0 = 1/2$ and thus, from Eq.~\ref{select}, an excitation  reduced by a factor of 4 as compared to at $M_1$. Linearly polarized light parallel to the crystal momentum of one of the M points, and with frequency tuned to the M point gap, thus generates significant excitation at only at that M point.

\begin{figure}[t!]
\centering\includegraphics[width=0.8\textwidth]{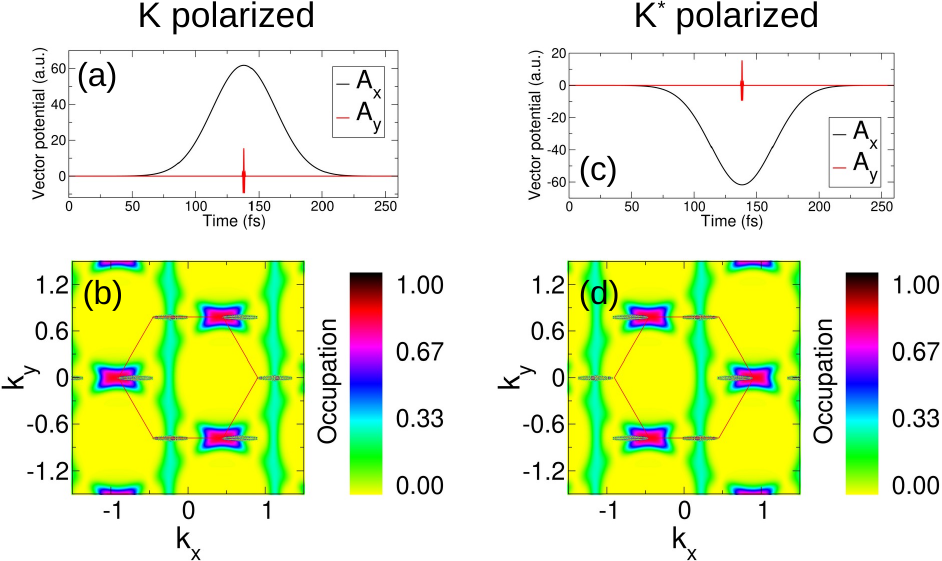}
\caption{\it{Valley polarization by double pumped light.} By combining a deep ultraviolet pulse tuned to the M point gap with an orthogonal THz pulse whose amplitude satisfies $A/c = |{\bf K} - {\bf M_1}|$, vector potential shown in panel (a), the charge excitation created at $M_1$ by the deep ultraviolet pulse is shifted to one of the K valleys, panel (b), in which is shown the momentum resolved excitation after the pulse. The deep ultraviolet pulse is the same as that employed in Fig.~\ref{fig1} and comparison of the momentum resolved excitations between the deep ultraviolet pulse acting alone, Fig.~\ref{fig1}(d), and in combination with a THz envelope, panel (b), clearly shows the change induced by the THz component. Switching the sign of the THz pulse amplitude selects for either polarization at the K valley, panels (a,b), or the K$^\ast$ valley, panels (c,d).
}
\label{fig2}
\end{figure}

We now consider the ultrafast real time excitation of charge by such a pulse. A few cycle linearly polarized pulse, with polarization vector aligned to $M_1$ and frequency of 4~eV, i.e. tuned to excite at $M_1$ according to the selection rule Eq.~\ref{select}, is applied to single layer graphene resulting in a pronounced charge excitation, Figs.~\ref{fig1}(a-c). The momentum resolved excited state charge at $t=5$~fs, Fig.~\ref{fig1}d, reveals striking inhomogeneity, with pronounced excitation at $M_1$, and much weaker and equal excitation at $M_{2,3}$, exactly as expected on the basis of the selection rule for M point light-matter coupling, Eq.~\ref{select}. One can note, however, the significant momentum space spread of the excitation that follows from the broadband nature of the ultrashort pulse employed.

A second pulse component is required to displace this localized charge excitation from the M point to the low energy K point of interest. To this end we employ the "hencomb" concept, recently invoked to create spin currents in WSe$_2$\cite{sharma23}, nearly pure valley current in minimally (40~meV) gapped graphene\cite{sharma_giant_2023}, and valley charge polarization in gapless graphene\cite{sharma_combining_2025}. A THz pulse of amplitude $(\v K - \v M_1)/c$ generates intraband evolution of crystal momentum from $\v K$ to $\v M_1$ and back to $\v K$; in contrast to the previously considered ultrafast M point excitation pulse, we take here a long duration pulse of full width half maxima 40~fs, see Fig.~\ref{fig2}(a,b). By acting at half-cycle of this THz pulse with the ultraviolet pulse we then have the situation in which intraband evolution evolves crystal momentum from $\v K$ to $\v M_1$ in the first half cycle, excitation occurs to the conduction band at $\v M_1$, but then this charge is evolved back from $\v M_1$ to $\v K$ in the second half cycle. The overall effect of this pulse is therefore to excite at the low energy K valley the charge that was, without the THz pulse, excited at the high energy M saddle point. Reversing the sign of the THz amplitude will, by similar argument, displace the excitation to the K$^\ast$ valley. The resulting momentum resolved excitations, shown in Fig.~\ref{fig2}(c,d) confirm exactly this picture, showing a distinct localized charge excitation at the K and K$^\ast$ valleys.

\section{Control over valley polarization}

Having established the basic effect of K valley polarization via the M point saddle, we now explore how this K valley polarization may be optimized. To this end a figure of merit is required, for which we employ the valley charge polarization defined as

\begin{equation}
\eta = \frac{Q_K - Q_{K^\ast}}{Q_K + Q_{K^\ast}}
\label{vp}
\end{equation}
where $Q_K$ is the charge exited at the K valley and $Q_{K^\ast}$ the charge excited at the conjugate K$^\ast$ valley. This parameter takes on the value of $\eta = +1$ in the case of a perfectly valley pure charge state, falling to $\eta = 0$ for valley uncontrasted excitation in which $Q_K = Q_{K^\ast}$. Note that a modified definition in which the denominator is replaced by the average valley charge, $(Q_K + Q_{K^\ast})/2$, is sometimes employed in the literature\cite{mrudul_light-induced_2021,
mrudul_controlling_2021}. The valley charges are obtained via integration of the excited charge in a zone corresponding to the Dirac-Weyl cone of each valley -- these are the circular domains shown in Fig.~\ref{fig1}(a).

\begin{figure}[t!]
\centering\includegraphics[width=1.0\textwidth]{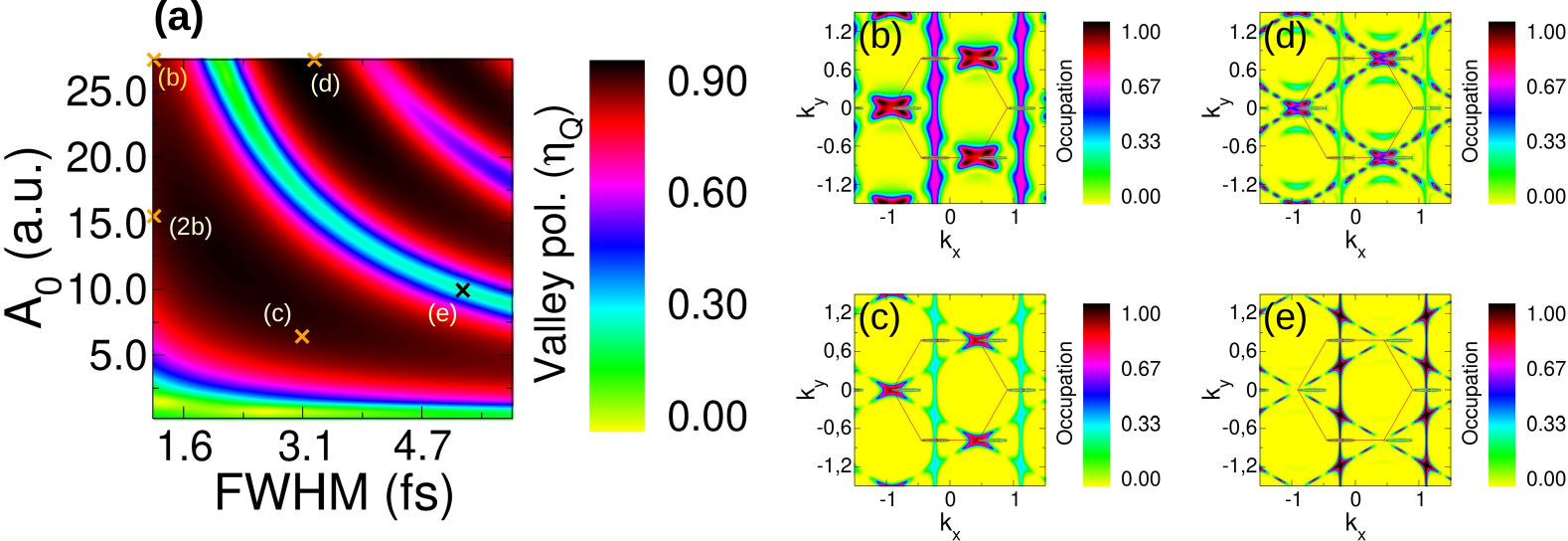}
\caption{\it{Optimizing valley polarization via the deep ultraviolet pulse component.} (a) The valley polarization of graphene created by the dual frequency deep ultraviolet and THz pulse, shown as a function of the full width half maxima (FWHM) and amplitude ($A_0$) of the deep ultraviolet component. The THz component is held fixed to the form shown in Fig.~\ref{fig2}(a). Regions of strong valley polarization can be seen to alternate with regions in which the valley polarization falls nearly to zero. Panels (b-e) display the momentum resolved excitation for four representative cases, as labelled in panel (a). The pulse parameters corresponding the Fig.~\ref{fig2}(a,b) are also indicated as the point "(2b)".
}
\label{fig3}
\end{figure}

The amplitude of the THz envelope is fixed by the requirement to displace the saddle charge excitation to the low energy valley, $\v A_0 = (\v K - \v M_1)/c$, while the frequency of the ultraviolet component is fixed by the gap at the M point. This leaves the duration and amplitude of the ultraviolet component, and the duration of the THz component, as pulse freedoms. We will now explore each of these in turn.

\begin{figure}[t!]
\centering\includegraphics[width=0.9\textwidth]{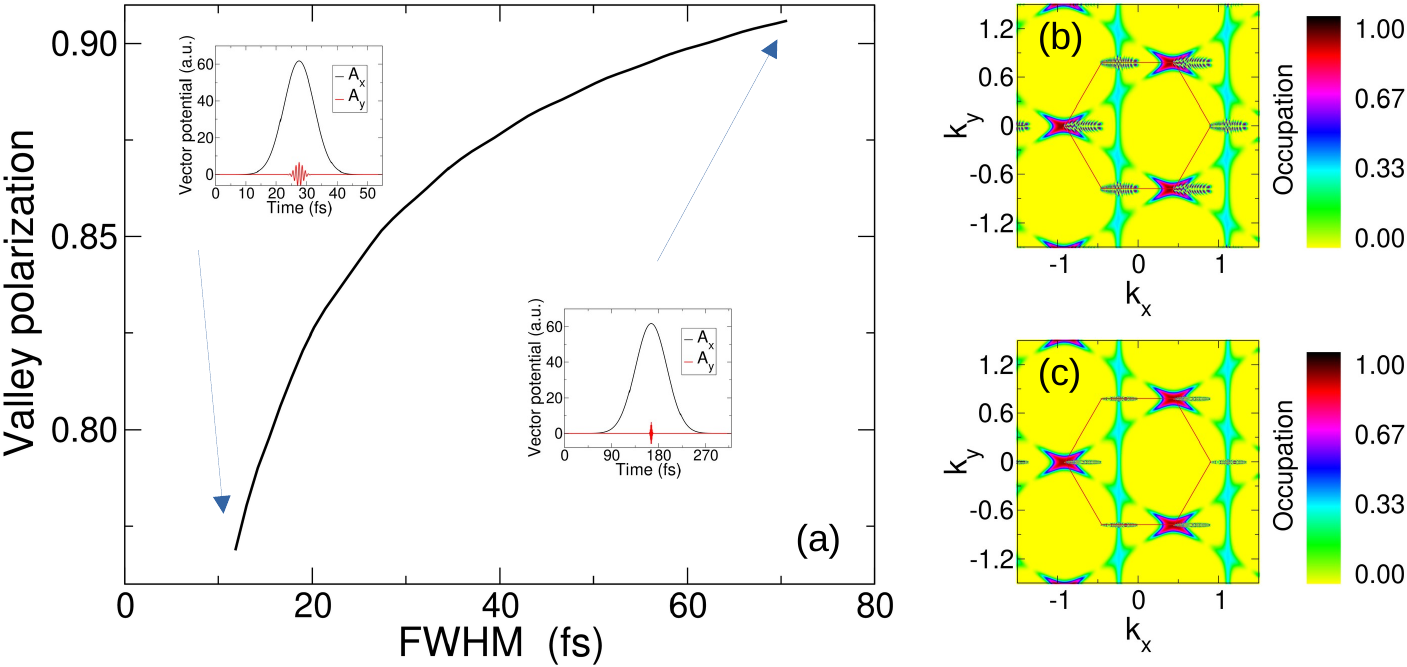}
\caption{\it{Variation of the valley polarization as a function of the duration of the THz pulse component.} The charge excited by the THz component -- two lines of charge in the K-M direction at each valley -- depends on the THz electric field which, as the amplitude of this pulse component is fixed to $\v A_0 = (\v K-\v M_1)/c$, depends on the pulse duration. Reduction of the THz full width half maximum (FWHM) increases the THz electric field and excited charge, lowering the valley polarization, as shown in panel (a). The momentum resolved charge excitation for two representative pulses at the temporal limits of the pulse envelope, as labelled in panel (a), are shown in panels (b) and (c).
}
\label{fig4}
\end{figure}

In Fig.~\ref{fig3}(a) we present a "light pulse map" of the valley polarization created by varying the FWHM and amplitude of the deep ultraviolet pulse component. A distinct series of bands of high and low valley polarization can be seen. By tuning the pulse FWHM and amplitude the valley polarization can be optimized to values of nearly $\eta = 0.9$.

The "bands" of valley low polarization would appear to represent a case in which the saddle selection rule has broken down. To probe this we present in panels (b-e) the momentum resolved excitations corresponding to four representative cases, as labelled in Fig.~\ref{fig3}(a). Panel (e) represents the situation in which a low valley polarization is achieved by the pulse, and as can be seen the charge excitation here is shifted off the K valley. This occurs due to Rabi oscillations such that the charge at $M_1$ de-excites with a corresponding increase in charge at the two partner saddles $M_2$ and $M_3$. As the THz pulse by design connects only the $M_1$ saddle to the K valley, as the THz pulse amplitude is given by $\v A_0 = (\v K-\v M_1)/c$, then strong excitation at the $M_2$ or $M_3$ saddles results in the pulse generating a charge excitation displaced from the K valley.

The THz pulse duration represents a final "control knob" that can be tuned to optimize valley polarization. In Fig.~\ref{fig4} we see that increasing the THz pulse duration generates a monotonically increasing valley polarization. This arises from the fact that as graphene is gapless, any light pulse that generates an intraband evolution of momentum intersecting the Dirac point will generate charge excitation. The THz pulse thus itself generates a line of excited charge extending from each K valley, as may be seen Fig.~\ref{fig2} and FIg.~\ref{fig3}. Evidently, increasing the THz pulse duration at fixed amplitude will reduce the electric field and thus reduce this valley uncontrasted charge excitation, exactly as seen in Fig.~\ref{fig4}. The limiting cases of large and small THz duration, whose corresponding momentum resolved excitations as shown in panels (b) and (c), clearly show the reduction in the THz excited lines of charge emanating from the K valleys upon increased duration of the THz pulse.

\section{{\it Ab-initio} calculation of valley polarization}

Having considered a mechanism of valley polarization via the M point on the basis of a tight-binding model, we now consider this physics in the context of {\it ab-initio} full potential time-dependent density functional theory calculations (TD-DFT).

In a recent work Sharma {\it et al.} reported a significant qualitative difference between the charge dynamics as generated by the tight-binding method, and by the more accurate TD-DFT method\cite{sharma_combining_2025}. In a similar scheme to that presented here, a linearly polarized THz pulse was employed to shift the K and K$^\ast$ charge excitation generated by an infra-red circularly polarized pulse to the K and $\Gamma$ points. This pulse would, therefore, be expected to yield a very high valley polarization -- as charge is excited at only one of the valleys -- along with a high energy $\Gamma$ point excitation. In a tight-binding simulation this was exactly what was seen, however a TD-DFT simulation revealed significant de-excitation as the THz pulse evolved charge from the K valley to the $\Gamma$ point, with the remarkable result that at the end of the pulse {\it only} the low energy K$^\ast$ valley charge remained.

\begin{figure}[t!]
\centering\includegraphics[width=1.0\textwidth]{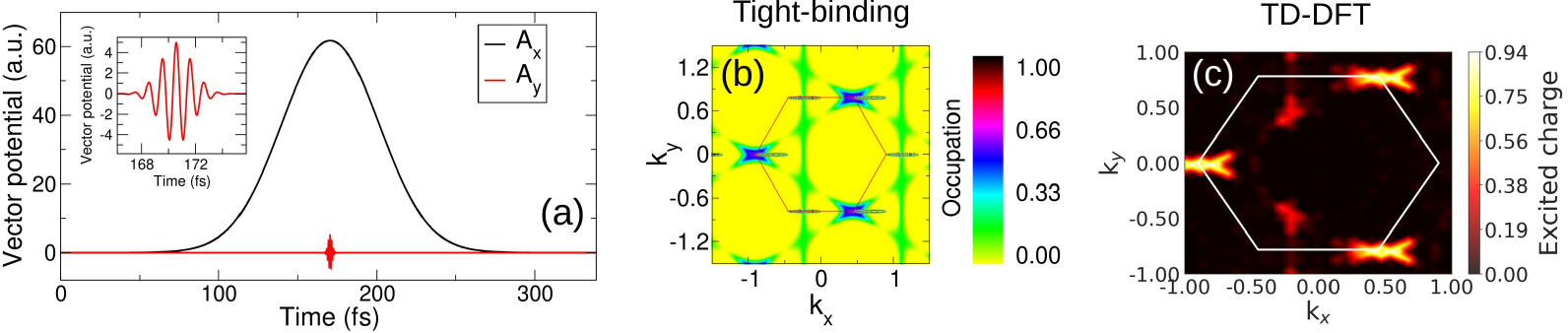}
\caption{\it{Comparison between (i) tight-binding and (ii) time dependent density functional theory for the charge excitation induced by double pumped THz and deep ultraviolet light in gapless graphene.} (a) Vector potential of a dual frequency laser pulse combining orthogonally polarized THz light (the Gaussian envelope) and a deep ultraviolet pulse tuned to the M point gap (shown also as the inset panel). (b,c) The charge excitation after the pulse calculated, respectively, with the tight-binding and time dependent density functional theory. While the principal result of charge excitation at the K$^\ast$ valley can be seen in both cases, the charge excitation generated by the THz component at the K valley -- clearly visible in the TB result -- is absent in the TD-DFT simulation.
}
\label{fig5}
\end{figure}

To explore whether this occurs also in the present case we consider the dynamics induced by the pulse shown in Fig.~\ref{fig5}a, comparing the resulting charge excitation generated in the tight-binding and TD-DFT methods. In panels (b) and (c) are shown the momentum resolved excitation after the pulse in the tight-binding and TD-DFT methods respectively. A comparison between these reveals that the "line" of charge excited by the THz pulse at the K valley -- clearly visible in panel (b) -- is missing in the TD-DFT calculation shown in panel (c). This suggests that charge de-excitation has occurred, such that when the THz pulse evolves intraband momentum from some point on the K-M$_1$ line to intersect with K -- thus exciting charge into the conduction band -- this charge is then de-excited as the second half cycle of the pulse evolves the intra-band momentum back to the initial crystal momentum. The result, as can be seen in panel (c), is a near perfect low energy valley polarization.

In previous work we have not observed any {\it qualitative difference} between the dynamics as simulated in TB and TD-DFT\cite{
sharma_valley_2022,
sharma23,
sharma_direct_2024,
gill_ultrafast_2025}; the de-excitation observed in the present work, with a similar result reported in Ref.~\cite{sharma_combining_2025}, is however clear evidence of a qualitatively different dynamics. Evidently, full TDDFT allows a THz de-excitation mechanism in graphene that is not found in the simpler tight-binding approach. For generating charge polarized states in graphene this is useful, as excitation by linearly polarized THz light does not distinguish between valleys and so the de-excitation of such charge increases valley contrast. This, furthermore, suggests that previous tight-binding based calculations in the literature\cite{kelardeh_ultrashort_2022,
avetissian_graphene_2023} involving THz pulses and graphene may in fact yield a better valley polarizations than has been reported.

\section{Discussion}

Valley polarization of graphene cannot proceed on the basis of the selection rule that governs the polarization of valley states in gapped materials, such as the $2H$ transition metal dichalcogenides. To solve the problem of generating a valley polarized charge excitation in graphene, previous work has exploited  the valley $C_3$ symmetry and the mirror relation between the K and K$^\ast$ manifolds.

The {\it local} version of this is the "trigonal warping" of the valley manifolds, which can be utilized by a pulse whose Lissajous figure has (approximate) $C_3$ symmetry. By designing such a pulse so that the three points of peak electric field are coincident with the maximum valence-conduction separation at one valley, and the minimum separation at its conjugate partner, valley contrast can be achieved. This attractive idea, however, generates only weak valley polarizations of $\eta = 0.1-0.2$ (employing Eq.~\ref{vp} for the definition of $\eta$) \cite{
mrudul_light-induced_2021,
mrudul_controlling_2021}.

A second approach is to utilize the {\it global} valley symmetry via a pulse that evolves the intraband momentum from K to K$^\ast$\cite{kelardeh_ultrashort_2022,
avetissian_graphene_2023,sharma_combining_2025}. At K this will evolve momentum from K to K$^\ast$ and back, while at the partner K$^\ast$ valley, due to the mirror symmetry relation, the momentum evolution will be from K$^\ast$ to $\Gamma$ and back. As charge cannot be excited at $\Gamma$, this implies contrast between the excitation that will occur at the two valleys. This idea has been exploited both for linearly polarized light\cite{avetissian_graphene_2023}, as well as via a "pedestal" pulse\cite{kelardeh_ultrashort_2022}. Finally, in the "hencomb" implementation of this idea, very high valley charge polarization states of $\eta=0.9$ can be achieved\cite{sharma_combining_2025}.

In the present approach, we have, in contrast, employed a selection rule valid at the M points of graphene and, by augmenting the light pulse by a THz envelope, imparted a momentum shift displacing the excitation from the M point saddle to the low-energy K valley. Alteration of the sign of the THz pulse amplitude switches the excitation between the K and K$^\ast$ valleys. The M point selection rule is thus by THz light "recast" as a route towards control over low-energy valley polarization. This avoids the large electric fields required to evolve momentum between the K and $\Gamma$ points. The resulting charge excitation we have shown to be both highly localized at the low-energy K valleys and to exhibit high valley polarization. Although we have considered only single-layer graphene, this approach should work equally well in other gapless members of the Xene family and in few-layer graphenes such as twist bilayer graphene. Our method thus opens a route to the lightwave control of valley physics in gapless materials via the saddle point.

\backmatter


\bmhead{Ethics, Consent to Participate, and Consent to Publish declarations}

Not applicable.

\bmhead{Funding Declaration}

Gill would like to thank DFG for funding through project-ID 328545488 TRR227 (project A04) Sharma would like to thank DFG for funding through project-ID 328545488 TRR227 (projects A04), and Shallcross would like to thank DFG for funding through project-ID 522036409 SH 498/7-1.

\bmhead{Clinical trial number}

Not applicable.

\bmhead{Acknowledgements}

Sharma would like to thank DFG for funding through project-ID 328545488 TRR227 (projects A04). The authors acknowledge the North-German Supercomputing Alliance (HLRN) for providing HPC resources that have contributed to the research results reported in this paper.

\bmhead{Author contributions}

The project was designed by SShall and P.E, D.G and P.E performed the \emph{ab-initio} calculations, SShar and K.D wrote the relevant code, all authors contribution to discussions and writing of the manuscript.

\bmhead{Data availability}

All data associated with this manuscript is available upon reasonable request.


\end{document}